\documentclass[prx, 10pt, twocolumn, floafix, superscriptaddress]{revtex4-1}

\usepackage{times, mathrsfs, amsmath, amsfonts, ntheorem, graphics, graphicx, color, theorem, bbm, mathtools, amssymb, xr}

\usepackage[english]{babel}

\usepackage[unicode=true, breaklinks=false, pdfborder={0 0 1}, backref=false, colorlinks=true, linkcolor=blue, citecolor=blue]{hyperref}

\newcommand{\cptp}{\textsc{CPTP}}
\newcommand{\cp}{\textsc{CP}}

\newcommand{\tr}{\mbox{tr}}

\def\braket#1{\mathinner{\langle{#1}\rangle}}

\def\BraVert{\egroup\,\mid\,\bgroup}

\newtheorem*{definition*}{Definition}

\newtheorem*{proposition*}{Proposition}
\newtheorem*{lemma*}{Lemma}
\newtheorem*{algorithm*}{Algorithm}
\newtheorem*{fact*}{Fact}
\newtheorem*{theorem*}{Theorem}
\newtheorem*{corollary*}{Corollary}
\newtheorem*{conjecture*}{Conjecture}
\newtheorem*{postulate*}{Postulate}
\newtheorem*{axiom*}{Axiom}
\newtheorem*{remark*}{Remark}
\newtheorem*{example*}{Example}
\newtheorem*{question*}{Question}

\begin{document} 

\title{Operational Markov condition for quantum processes}

\author{Felix A. Pollock}
\email{felix.pollock@monash.edu}
\affiliation{School of Physics \& Astronomy, Monash University, Clayton, Victoria 3800, Australia}

\author{C\'esar Rodr\'iguez-Rosario}
\affiliation{Bremen Center for Computational Materials Science, University of Bremen, Am Fallturm 1, D-28359, Bremen, Germany}

\author{Thomas Frauenheim}
\affiliation{Bremen Center for Computational Materials Science, University of Bremen, Am Fallturm 1, D-28359, Bremen, Germany}

\author{Mauro Paternostro}
\affiliation{School of Mathematics and Physics, Queen’s University, Belfast BT7 1NN, United Kingdom}

\author{Kavan Modi}
\email{kavan.modi@monash.edu}
\affiliation{School of Physics \& Astronomy, Monash University, Clayton, Victoria 3800, Australia}

\date{\today}

\begin{abstract}
We derive a necessary and sufficient condition for a quantum process to be Markovian which coincides with the classical one in the relevant limit. Our condition unifies all previously known definitions for quantum Markov processes by accounting for all potentially detectable memory effects. We then derive a family of  measures of non-Markovianity with clear operational interpretations, such as the size of the memory required to simulate a process, or the experimental falsifiability of a Markovian hypothesis.
\end{abstract}
\maketitle 

In classical probability theory, a stochastic process is the collection of joint probability distributions of a system's state (described by random variable $X$) at different times, $\{P(X_k,t_k; X_{k-1},t_{k-1};\dots; X_1,t_1; X_0,t_0) \,\forall k \in \mathbb{N}\}$; to be a valid process, these distributions must additionally satisfy the Kolmogorov consistency conditions~\cite{BreuerPetruccione}. A Markov process is one where the state $X_k$ of the system at any time $t_k$ only depends conditionally on the state of the system at the previous time step, and not on the remaining history. That is, the conditional probability distributions satisfy
\begin{gather}\label{eqn::Cl-Markov}
\!P(X_k,t_k|X_{k\!-\!1},t_{k\!-\!1};\!\dots\!;X_0,t_0) \!=\! P(X_k,t_k|X_{k\!-\!1},t_{k\!-\!1})
\end{gather}
for all $k$. This simple looking condition has profound implications, leading to a massively simplified description of the stochastic process. The study of such processes forms an entire branch of mathematics, and the evolution of physical systems is frequently approximated to be Markov (when it is not exactly so). This is in part due to the fact that the properties of Markov processes make them easier to manipulate analytically and computationally~\cite{van2011stochastic}. 

Implicit in this description of a classical process is the assumption that the value of $X_j$ at a given time can be observed without affecting the subsequent evolution. This assumption cannot be valid for quantum processes. In quantum theory, a measurement must be performed to infer the state of system. And the measurement process, in general, \emph{must} disturb that state. Therefore, unlike its classical counterpart, a generic quantum stochastic process cannot be described without interfering with it~\footnote{Allowing for interventions in classical setting leads to a much richer theory~\cite{Pearl}, where correlations and causation can be differentiated.}. These complications make it challenging to define the process independently of the control operations of the experimenter. From a technical perspective, a serious consequence of this is that joint probability distributions of quantum observables at different times do not satisfy the Kolmogorov conditions~\cite{BreuerPetruccione}, and do not constitute stochastic processes in the classical sense.

Nevertheless, temporal correlations between observables do play an important role in the dynamics of many open quantum systems, e.g. in the emission spectra of quantum dots~\cite{mccutcheon2016} and in the vibrational motion of interacting molecular fluids~\cite{tanimura2015}. Quantifying memory effects, and clearly defining the boundary between Markovian and non-Markovian quantum processes, represents an important challenge in describing such systems.

Attempts at solving this problem tend to take a \emph{necessary}, but not sufficient, condition for a classical process to satisfy Eq.~\eqref{eqn::Cl-Markov}, and extend it to the quantum domain. This has led to a zoo of quantum Markov definitions, and accompanying ``measures'' of non-Markovianity~\cite{NMrev, RevModPhys.88.021002}, that do not coincide with Eq.~\eqref{eqn::Cl-Markov} in the classical case~\footnote{Some measures claim to be based on \textit{necessary and sufficient} Markov conditions, but only with respect to a quantum Markov definition that does not reduce to the classical one in the correct limit.}. Examples include measures based on: monotonicity of trace-distance distinguishability~\cite{PhysRevLett.103.210401}; the divisibility of dynamics~\cite{PhysRevLett.105.050403, hou2011}; how quantum Fisher information changes~\cite{PhysRevA.82.042103}; the detection of initial correlations~\cite{mazzola2012dynamical, rodriguez2012unification, rodriguez2008completely, modi_operational_2012, IC-breuer, rodriguez2012unification, PhysRevLett.107.180402, Gessner:2014kl}; changes to quantum correlations or coherence~\cite{PhysRevA.86.044101, zhihe2017}; channel capacities and information flow~\cite{fanchini2014, bylicka2014, pineda2016, bylicka2016}; and positivity of quantum maps~\cite{PhysRevA.83.022109, PhysRevLett.101.150402, rajagopal, sabri}.

All these methods offer valid ways to witness memory effects. Unfortunately, however, they often lack a clear operational basis. Moreover, different measures of non-Markovianity agree neither on the degree of non-Markovianity of a given process, nor even on whether it is Markovian~\cite{PhysRevA.83.052128}. Put another way, they each fail to quantify demonstrable memory effects in some cases. These inconsistencies have led some to the conclusion that there can be no unique condition for a quantum Markov process. 

In this Letter, we use the process tensor framework, introduced in an accompanying article~\cite{PRA_partner}, to demonstrate that this conclusion is false. We first present a robust operational definition for a quantum Markov process, which unifies all previous definitions and, most importantly, reduces to Eq.~\eqref{eqn::Cl-Markov} for classical processes. We then go on to derive a family of measures for non-Markovianity which quantify \emph{all} detectable memory effects, and which have a clear operational interpretation. 

\textit{Quantum stochastic processes}---Conventional approaches to open quantum dynamics describe a process solely in terms a system's time-evolving density matrix $\rho_t$, which is related to the initial state of the system by a completely positive trace-preserving (\textbf{\cptp}) map $\Lambda_{t:0}$. However, as has also been argued in the classical case~\cite{vanKampen1998}, a framework that captures non-Markovian effects cannot be a simple extension of one which characterises memoryless processes. In order to describe the joint probability distributions of multiple measurement outcomes, and hence capture memory effects which only appear in multi-time correlation functions, we must go beyond the paradigm of \cptp~maps~\footnote{It is not enough to simply relinquish the complete positivity of the dynamics, as one might think following the arguments of Pechukas and Alicki~\cite{pechukas, alicki, pechukas2}. The not completely positive maps formalism is not operationally consistent, and cannot be used to determine multi-time correlation functions~\cite{arxiv:1708.00769}.}.

We consider a scenario where the role of the observer in a stochastic process is made explicit: A series of control operations $\mathcal{A}_j^{(r)}$ act on the system at times $t_j$ (here, $r$ labels one of a set of operations that could have been realised, with some probability, at that time). These can correspond to measurements, unitary transformations, interactions with an ancilla or anything in between, and are represented mathematically by completely positive (\textbf{\cp}) maps. As implied above, their action need not be deterministic (for example, in the case of different measurement outcomes), but the average control operation applied at a given point corresponds to a deterministic \cptp~map $\mathcal{A}_j=\sum_r\mathcal{A}_j^{(r)}$. The choice of \cptp~map and its decomposition into operations $\mathcal{A}_j^{(r)}$ is often referred to as an \emph{instrument}, and the latter can equivalently be thought of as a decomposition of $\mathcal{A}_j$ into Kraus operators. The entire sequence of control operations at times $\{t_0,t_1,...t_{k-1}\}$ may, furthermore, be correlated, and we denote it by $\mathbf{A}_{k-1:0}$ (which is an element of the tensor product of spaces of control operations at each step). When the operations are uncorrelated, this can simply be thought of as the sequence $\mathbf{A}_{k-1:0}=\{\mathcal{A}^{(r_{k-1})}_{k-1};...;\mathcal{A}^{(r_1)}_1,\mathcal{A}^{(r_0)}_0\}$.

In an accompanying Article~\cite{PRA_partner}, we describe how a process can be fully characterised by a linear and \cp~mapping $\mathcal{T}_{k:0}$, called the process tensor, which takes a sequence of operations to the density operator at a later time: $\rho_k = \mathcal{T}_{k:0} [\mathbf{A}_{k-1:0}]$. $\mathcal{T}_{k:0}$ encodes all uncontrollable properties of the process, including any interactions of the system with its environment, as well as their (possibly correlated) average initial state. When the control operations are non-deterministic, $\rho_k$ is subnormalised, with a trace that gives the joint probability of applying those operations. Any given process tensor is guaranteed to be consistent with unitary dynamics of the system with a suitable environment.  If the process tensor, defined on any set of time steps in an interval, and the control operations all act in a fixed basis, then the description reduces to that of a classical stochastic process as described in the introduction. Interestingly, quantum stochastic processes have been defined in a mathematically related way several times in the past~\cite{lindblad_non-markovian_1979, accardi, kretschmann_quantum_2005}, without being widely adopted by the open quantum systems community.

Our description, in terms of the process tensor, fully contains the conventional one; doing nothing to the system, represented by the identity map $\mathcal{I}$ is a perfectly valid control operation and, for a system initially uncorrelated with its environment, $\mathcal{T}_{k:0} [\mathcal{I}^{\otimes k}] = \Lambda_{k:0}[\rho_0]$. The main achievement of the process tensor framework is to separate `the process,' as dictated by Nature, from an experimenter's control operations. In other words, the process tensor describes everything that is independent of the choices of the experimenter. Using this framework, we are now in a position to present our main result.

\emph{Criterion for a quantum Markov process.---} To clearly {and operationally} formulate a quantum Markov condition, we introduce the idea of a \emph{causal break}, where the system's state is actively reset, dividing its evolution into two causally disconnected segments. We then test for conditional dependence of the future dynamics on the past control operations. If the future process depends on the past controls, then we must conclude that the process carries memory and it is non-Markovian.

To formalise this notion, we begin by explicitly denoting the state of the system at time step $l$ as a function of previous control operations, $\rho_l = \rho_l (\mathbf{A}_{l-1:0})$. Now, suppose at time step $k < l$ we make a measurement (of our choice) on the system and observe outcome $r$, which occurs with probability $p_{k}^{(r)}$; the corresponding positive operator is denoted $\Pi^{(r)}_{k}$. We then re-prepare the system into a known state $P_{k}^{(s)}$, chosen randomly from some set $\{P_{k}^{(s)}\}$. The measurement and the re-preparation at $k$ break the causal link between the past $j \le k$ and the future $l > k$ of the system; more generally, any operation whose output is independent of its input constitutes a causal break. If we let the system evolve to time step $l$, its state will depend on the choice and the outcome of the measurement at $k$, the preparation $P_{k}$, and the control operations from $0$ to $k-1$. Therefore, we have a conditional subnormalised state $\tilde\rho_l= p_{r} \rho_l(P_{k}^{(s)}| \Pi^{(r)}_{k}; \mathbf{A}_{k-2:0})$, where the conditioning argument is the choice of past measurement $\Pi^{(r)}_{k}$ and controls $\{\mathbf{A}_{k-1:0}\}$. The probability $p_{r}$, which also, in general, depends on $\{\mathbf{A}_{k-1:0}\}$, is not relevant to whether the process is Markovian; we are interested only in whether the normalised state $\rho_l= \rho_l(P_{k}^{(s)}| \Pi^{(r)}_{k}; \mathbf{A}_{k-1:0})$ depends on its conditioning argument. This operationally well defined conditional state is fully consistent with conditional classical probability distributions. However, it is very different from the quantum conditional states defined in Ref.~\cite{arxiv:1607.03637}.

\begin{figure}[t]
\begin{center}
\includegraphics[width=0.85 \linewidth ]{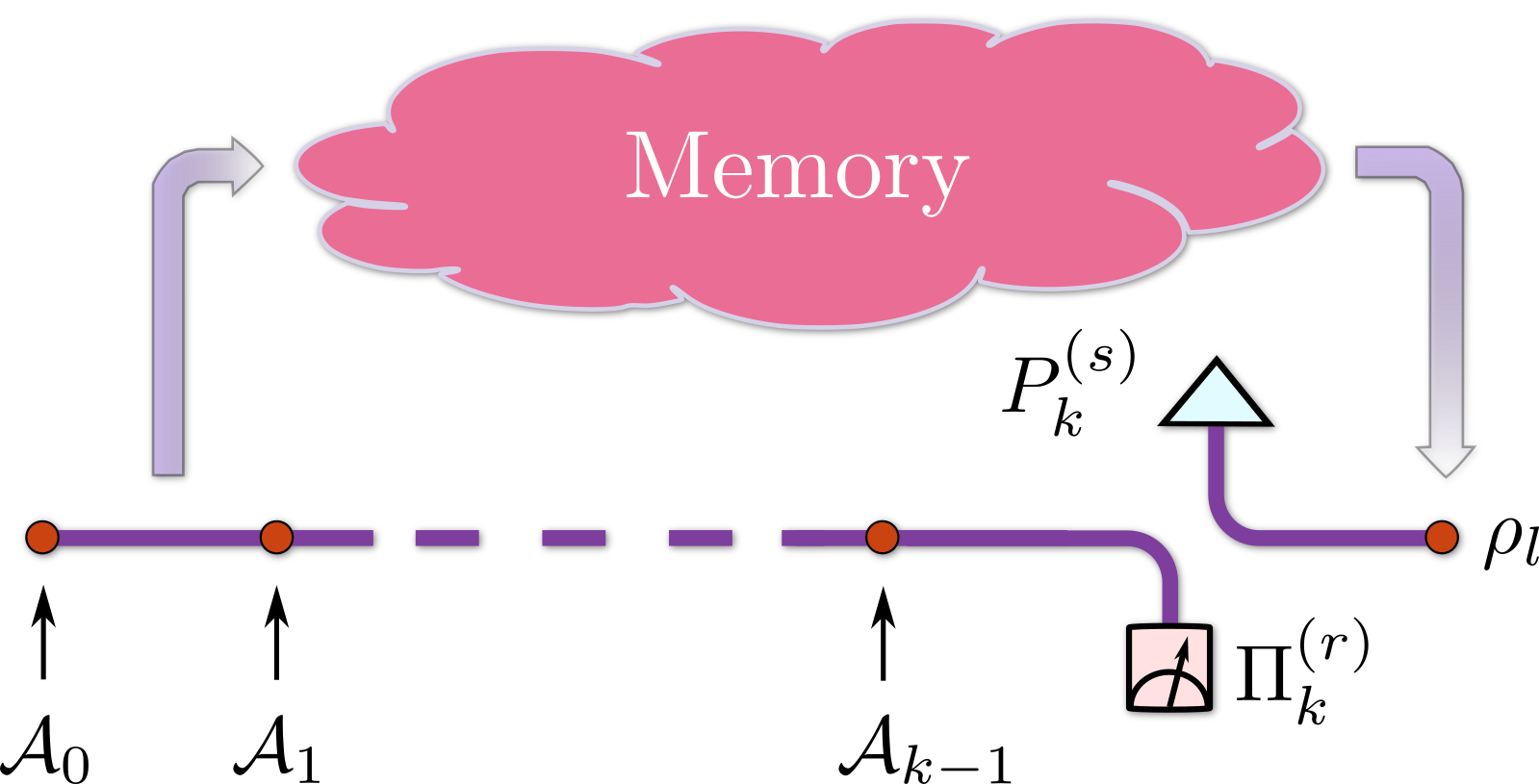}
\caption{\emph{Determining whether a quantum process is Markovian.} Generalised operations $\mathbf{A}_{k:0}$ are made on the system during a quantum process, where the subscripts represent the time. At time step $k$ we make a causal break by measuring the system with $\Pi^{(r)}_{k}$ and re-preparing it in randomly chosen state $P_{k}^{(s)}$. The process is said to be Markovian if and only if $\rho_l(P_{k}| \Pi^{(r)}_{k}; \mathbf{A}_{k-1:0}) = \rho_l(P_{k}^{(s)})$ at all time steps $l,k$, for all inputs $P_{k}^{(s)}$, measurements $\{ \Pi^{(r)}_{k}\}$, and control operations $\{\mathbf{A}_{k-1:0} \}$. \label{image-cloud}}
\end{center}
\vspace{-20pt}
\end{figure}

Because of the causal break, the system  itself cannot carry any information beyond step $k$ about $\Pi^{(r)}_{k}$ or its earlier history. The only way $\rho_l$ could depend on the controls is if the information from the past is carried across the causal break via some external environment (see Appendix~\ref{app:ex} for some examples). We have depicted this in Fig.~\ref{image-cloud}, with the memory as a cloud that transmits information from the past to the future across the causal break. This immediately results in the following operational criterion for a Markov process:
\begin{definition*}\label{markovdef}
A quantum process is Markovian when the state of the system $\rho_l$, after a causal break at time step $k$ (with $l>k$), only depends on the input state $P_{k}^{(s)}$: $\rho_{l}(P_{k}^{(s)} |\Pi^{(r)}_{k};\mathbf{A}_{k-1:0}) = \rho_{l}(P_{k}^{(s)})$, $\forall \, \{P_{k}^{(s)},\Pi^{(r)}_{k},\mathbf{A}_{k-1:0}\}$ and $\forall \, l,\,k \in [0,K]$.
\end{definition*}

Note that this definition is directly analogous to the \emph{causal Markov condition} for a discrete-time classical stochastic evolution that allows for interventions~\cite{Pearl}: While the definition in Eq.~\eqref{eqn::Cl-Markov} refers only to the system state at different times, more modern descriptions of (classical) stochastic processes in terms of their causal structure allow for interventions between time steps. Recently, and independently of this work, a generalisation of this kind of `Markovian causal modelling' has been developed for quantum Markov processes~\cite{costashrapnel2016}.

From the Definition, we have the following Theorem:
\begin{theorem*}\label{thm:markov}
A quantum process is non-Markovian iff there exist at least two different choices of controls $\{\Pi^{(r)}_{k}; \mathbf{A}_{k-1;0}\}$ and $\{\Pi^{'(r')}_{k}; \mathbf{A}'_{k-1;0}\}$, such that after a causal break at time step $k$, the conditional states of the system at time step $l$ are different:
\begin{gather}\label{CLMarkovCond1}
\rho_{l}(P_{k}^{(s)}|\Pi^{(r)}_{k};\mathbf{A}_{k-1;0}) \ne \rho_{l}(P_{k}^{(s)}|\Pi^{'(r')}_{k};\mathbf{A}'_{k-1;0}).
\end{gather}
Conversely, if $\rho_{l}$ is constant for all linearly independent controls, then the process is Markovian.
\end{theorem*}
The proof, which relies on the linearity of the process tensor, is given in Appendix~\ref{app:markov}. Identifying two controls that lead to different conditional states may, in pathological cases, require testing Eq.~\eqref{CLMarkovCond1} for all possible (exponentially many) linearly independent control operations, though the discovery of any pair of control sequences that lead to an inequality in Eq.~\eqref{CLMarkovCond1} is a witness for non-Markovianity; this is directly analogous to the problem of testing for correlations in a many-body state. The implication of the Theorem is that it is possible to determine whether a process is Markovian in a finite number of experiments. 

Our Theorem also has the appealing consequence that quantum Markov processes give rise to classical ones:

\begin{corollary*}\label{cor:qincl}
Fixing a choice of instruments always leads to a classical probability distribution satisfying Eq.~\eqref{eqn::Cl-Markov} iff the quantum process is Markovian according the Definition provided above.
\end{corollary*}
\textit{Proof.} Fixing a choice of instruments means allowing only one of a set of operations $\mathcal{A}_{j}^{(r)}$ to act at each time step, such that $\sum_r \mathcal{A}_{j}^{(r)}$ is a \cptp~map (the instrument may be different at different time steps). As such, the trace of the state at time $k$ is the probability distribution $P(r_{k-1},t_{k-1};\dots;r_1,t_1;r_0,t_0) = \tr\rho_k(\mathcal{A}_{k-1}^{(r_{k-1})},\dots,\mathcal{A}_1^{(r_1)},\mathcal{A}_0^{(r_0)})$, where the $r_j$ can be treated as classical random variables. For a Markov process, we have that $\rho_j(\mathcal{A}_{j-1}^{(r_{j-1})},P_{j-2}^{(s)}|\Pi_{j-2}^{(r_{j-2})},\mathbf{A}_{j-3:0})=\rho_j(\mathcal{A}_{j-1}^{(r_{j-1})},P_{j-2}^{(s)}|\Pi_{j-2}^{(r_{j-2})})=\rho_j(\mathcal{A}_{j-1}^{(r_{j-1})}|P_{j-2}^{(s)},\Pi_{j-2}^{(s')})$ for any deterministic choice of preparation $P_{j-2}^{(s)}$. By writing $\mathcal{A}_{j-2}^{(r_{j-2})} = \sum_{s s'} c^{(r_{j-2})}_{s s'} P_{j-2}^{(s)}\otimes\Pi_{j-2}^{(s')}$~\footnote{Here we implicitly write the operation in terms of its Choi state.}, it follows that $P(r_{j-1},t_{j-1}|\dots;r_1,t_1;r_0,t_0) = P(r_{j-1},t_{j-1}|r_{j-2},t_{j-2})$~$\forall k>j>0$. From our Theorem, if the process is non-Markovian, then there is at least some pair of control operations for which the inequality in Eq.~\eqref{CLMarkovCond1} is true. By choosing an instrument which acts with these operations, one realises a classical process with $P(r_{j-1},t_{j-1}|r_{j-2},t_{j-2},\dots;r_0,t_0) \neq P(r_{j-1},t_{j-1}|r_{j-2},t_{j-2})$ for some values of $\{r_{j}\}$. \hfill $\blacksquare$

This remedies an important issue with existing definitions of quantum Markov processes; namely, that they fail to classify classical stochastic processes correctly~\cite{NMrev}. Instead, as discussed above, conventional approaches are based on necessary, but not sufficient, conditions for a classical process to be Markov. The above Corollary demonstrates that our Definition corresponds to a necessary and sufficient condition. Of course, those necessary conditions are still satisfied by Markov processes in our framework. In particular, we have the following Lemma:

\begin{lemma*}
\label{cor:divisible}
Markov processes are $K$-divisible, i.e., they can be written as a sequence of \textsc{cptp}~maps between the $K$ time steps on which they are defined.
\end{lemma*}
\textit{Proof.} If the condition introduced in our Definition is satisfied, then $\rho_k$ only depends on the previous choice of input $P_{k-1}^{(s)}$ for any $k$. By choosing from a complete set of linearly independent inputs $\{P_j^{(\nu_j)}\}$, quantum process tomography can be performed independently for each pair of adjacent time steps. Since the dynamics between any two time steps is free from the past (there is no conditioning on prior operations), the resulting set of \textsc{cptp}~maps completely describes the dynamics. These maps can then be composed to calculate the dynamics between any two time steps. In other words, the dynamics between time steps $l>k>j$ is described by maps $\Lambda_{k:j}$, $\Lambda_{l:k}$, and $\Lambda_{l:j}$, with the last map being the composition of the former two: $\Lambda_{l:j} = \Lambda_{l:k} \circ \Lambda_{k:j}$. \hfill $\blacksquare$

This means our result verifies the well-known hypothesis that Markovian dynamics is divisible. However, the converse of this statement does not hold, contrary to what is often postulated~\cite{NMrev}. That is, $\Lambda_{l:j} = \Lambda_{l:k} \circ \Lambda_{k:j}$ $\forall l>k>j \in [0,K]$ does not imply that the process is Markovian according to our main Theorem. In principle, there could be multi-time correlations between time steps that affect future dynamics \emph{conditioned} on past operations. In this light, the Theorem we present here can be seen as both a unification and generalisation of previous theories of quantum non-Markovianity, since all of these require non-Markovian processes to be indivisible. This direct consequence of the above Lemma is encapsulated in the following Remark:

\begin{remark*}
\label{rmk:unify}
Any process labelled non-Markovian according to the definitions given in Refs.~\cite{PhysRevLett.103.210401, PhysRevLett.105.050403, hou2011, PhysRevLett.105.050403, PhysRevA.82.042103, mazzola2012dynamical, rodriguez2008completely, modi_operational_2012, IC-breuer, rodriguez2012unification, PhysRevLett.107.180402, Gessner:2014kl, rodriguez2012unification, PhysRevA.86.044101, zhihe2017, fanchini2014, bylicka2014, pineda2016, bylicka2016, PhysRevA.83.022109, PhysRevLett.101.150402, rajagopal, sabri} will be non-Markovian according to our main  Theorem. The converse does not hold.
\end{remark*}

In fact, because it contains information about the density operator as a function of time, the process tensor formalism could be used to explicitly calculate any of the measures of non-Markovianity introduced in the above references. In Appendix~\ref{app:ex}, we give several examples of non-Markovian effects which are not detected by conventional approaches, but which are detected in our framework. The first manifests the discussion below the above Lemma, demonstrating that divisible (even \textsc{cp}-divisible) dynamics can have memory. We also show how the trace-distance definition of Markov processes can fail to characterise non-Markovianity, and that a quantum process can be non-Markovian even when there are no system-environment quantum correlations. 

It is worth noting that all open quantum evolutions generated by a time-independent system-environment Hamiltonian are non-Markovian according to our main Theorem, when considering more than two time steps. A similar point was also made in Ref.~\cite{PhysRevA.92.022102}, albeit in the context of dynamical decoupling. The strictness of the operational Markov Definition, however, does not render the notion of non-Markovianity meaningless; on the contrary, it allows us to construct meaningful measures of non-Markovianity.

\emph{Quantifying non-Markovianity.---}
One of the key features of the process tensor formalism is the isomorphism between a process $\mathcal{T}_{k:0}$ and a many-body generalised Choi state $\Upsilon_{k:0}$. The correlations between subsystems in $\Upsilon_{k:0}$ encode the temporal correlations in the corresponding process. As we prove in our Lemma above, a Markov process is divisible, i.e., it can be described by a sequence of independent \textsc{cptp}~maps. The corresponding Choi state will only have correlations between subsystems corresponding to neighbouring preparations and subsequent measurements; it can be written as the tensor product $\Upsilon_{k:0}^{\rm Markov} = \Lambda_{k:k-1}\otimes \Lambda_{k-1:k-2} \otimes \dots \otimes \Lambda_{1:0}\otimes \rho_0$, where $\Lambda_{j+1:j}$ is the Choi state of the \textsc{cptp}~map between time steps $j$ and $j+1$, and $\rho_0$ is the average initial state of the process. 

This observation allows us to define a degree of non-Markovianity.

\begin{proposition*}
\label{prop:nonmarkovmes}
Any \cp-contractive quasi-distance $\mathcal{D}$ between the generalised Choi state of a non-Markovian process and the closest Choi state of a Markov process measures the degree of non-Markovianity:
\begin{gather}\label{nonmarkovmes}
\mathcal{N} := \min_{\Upsilon_{k:0}^{\rm Markov}}\mathcal{D} \left[ \Upsilon_{k:0} \|\Upsilon_{k:0}^{\rm Markov} \right].
\end{gather}
\end{proposition*}
Here, \cp~contractive means that $\mathcal{D} [\Phi(X) \| \Phi(Y)] \leq \mathcal{D}[X\|Y]$ for any \cp~map $\Phi$ on the space of generalised Choi states, and a quasi-distance satisfies all the properties of a distance except that it may not be symmetric in its arguments. Other quasi-distance measures may also be used, with different operational interpretations, but those which are not CP-contractive do not lead to consistent measures for non-Markovianity~\footnote{Any measure based on a distance which is not CP contractive could be trivially decreased by the presence of an independent ancillary Markov process.}. 
If we choose relative entropy~\cite{vedral2002role} as the metric, then the closest Markov process is straightforwardly found by discarding the correlations. This measure of non-Markovianity has an operational interpretation: ${\rm Prob}_{\rm confusion} = \exp\{-n \mathcal{N}\}$ measures the probability of confusing the given non-Markovian process for a promised Markovian process after $n$ measurements of the Choi state. In other words,  $\Upsilon_{k:0}^{\rm Markov}$ represents a Markovian hypothesis for an experiment that is really described by $\Upsilon_{k:0}$. If $\mathcal{N}$ is large, then an experimenter will very quickly realise that the hypothesis is false, and the model needs updating.

Furthermore, other meaningful definitions of non-Markovianity can be derived from the properties of the Choi state. For example, the bond dimension of the matrix product representation of $\Upsilon_{k:0}$ indicates the size of the system required to store the memory between time steps; it is unity (no memory) only in the case of a Markov process. This clearly has importance for the efficiency of numerical simulations of complex quantum systems.

\emph{Discussion.---}
We have used the process tensor framework to introduce an unambiguous condition for quantum Markov dynamics. This condition is constructed in an entirely operational manner; and it meaningfully corresponds to the classical one in relevant settings. We have then used this condition to derive a family of measures for non-Markovianity, including one with a natural interpretation in terms of hypothesis testing with a Markovian model. Our measure will therefore enable experimenters to incrementally construct better models for a given system, by accounting for non-trivial non-Markovian memory. By means of the Trotter formula we can also extend the measure for non-Markovianity to continuous processes.

There are well-known methods to develop master equations for Markov processes. We can meaningfully quantify the error associated with using such methods for non-Markovian processes if we can bound their fidelity using Eq.~\eqref{nonmarkovmes}. This should be possible in many cases, since large environments tend not to retain long-term memory. We anticipate that most processes of physical interest will be almost Markovian and the corresponding process tensor should be highly sparse with a block-diagonal structure. In fact, equipped with a suitable measure on the space of Choi states, our Proposition allows for quantitative statements about typical non-Markovianity to be made, though we leave this for future work.

Because it captures all operationally accessible memory effects (and no more), the framework we have introduced in this Letter enables the unambiguous comparison of non-Markovianity between different systems. In particular, the fact that it puts quantum and classical processes on the same footing, will allow for a meaningful quantification of the advantages (or not) that quantum mechanics brings when using memory as a resource.

\begin{acknowledgments}
We are grateful to A. Aspuru-Guzik, G. Cohen, A. Gilchrist, J. Goold, M. W. Hall, T. Le, K. Li, L. Mazzola, S. Milz, F. Sakuldee, D. Terno, S. Vinjanampathy, H. Wiseman, C. Wood, M.-H. Yung for valuable conversations. CR-R  is supported by MSCA-IF-EF-ST - QFluctTrans 706890. MP is supported by the EU FP7 grant TherMiQ (Grant Agreement 618074), the DfE-SFI Investigator Programme (grant 15/IA/2864), the H2020 Collaborative Project TEQ (grant 766900), and the Royal Society. KM is supported through ARC FT160100073.
\end{acknowledgments}

\bibliography{Pollockbib}

\newpage
\appendix

\onecolumngrid

\section{Proof of quantum Markov condition (main Theorem)}\label{app:markov}

The first statement follows trivially from the definition of a quantum Markov process: if the inequality in Eq.~\eqref{CLMarkovCond1} holds, then the state at $l$ depends on the past beyond the input $P_{k}^{(s)}$.

We now proceed to prove the converse statement: if the left and right sides of Eq.~\eqref{CLMarkovCond1} are equal for a complete, linearly independent set of controls, they will be equal for any pair of controls, implying that the process is Markovian. First, consider expanding the process tensor for a general control sequence, prior to a causal break, in terms of the basis  $\{\mathcal{A}_j^{{(\mu,\nu)}_j}\}$: 
\begin{align}
\rho_{l}(P_{k}^{(s)},\Pi^{(r)}_{k};\mathbf{A}_{k-1:0})=& \mathcal{T}_{l:0} \left( P_{k}^{(s)}\otimes\Pi^{(r)}_{k},\mathbf{A}_{k-1:0} \right) \nonumber \\
=&\!\!\sum_{\vec{\mu},\vec{\nu},\mu_k}\!\! \alpha_{r,\mu_k}\alpha_{(\vec{\mu},\vec{\nu})} \, \rho_l \left( P_{k}^{(s)}\!\otimes\!\Pi^{(\mu_k)}_{k};\mathcal{A}_{k-1}^{{(\mu,\nu)}_{k-1}} ; \dots ; \mathcal{A}_{1}^{{(\mu,\nu)}_{1}} ; \mathcal{A}_{0}^{{(\mu,\nu)}_0} \!\right),\label{causalbreakdecomp}
\end{align}
where we are using the same notation as in Ref.~\cite{PRA_partner}. We have also expanded the POVM element $\Pi^{(r)}_{k}=\sum_{\mu_k}\alpha_{r,\mu_k}\Pi^{(\mu_k)}_{k}$ in terms of an informationally-complete set of basis POVM elements $\{\Pi^{(\mu_k)}_{k}\}$, and have assumed no further operations are applied between time steps $k$ and $l$ (the following proof straightforwardly generalizes to the case where later operations are applied). 

Using the definition in the main text, we can rewrite Eq.~\eqref{causalbreakdecomp} in terms of conditional states as
\begin{align}
\rho_{l}(P_{k}^{(s)},\Pi^{(r)}_{k};\mathbf{A}_{k-1:0})=& \sum_{\vec{\mu},\vec{\nu},\mu_k} \alpha_{r,\mu_k}\alpha_{(\vec{\mu},\vec{\nu})} p_{l}(\mu_k,\vec{\mu},\vec{\nu}) \nonumber \\
&\qquad\qquad\qquad\times\rho_l(P_{k}^{(s)}|\Pi^{(\mu_k)}_{k};\mathcal{A}_{k-1}^{{(\mu,\nu)}_{k-1}};\dots;\mathcal{A}_{1}^{{(\mu,\nu)}_{1}};\mathcal{A}_{0}^{{(\mu,\nu)}_0}),
\end{align}
where $p_{l}(\mu_k,\vec{\mu},\vec{\nu})$ is the joint probability distribution for the outcome corresponding to $\Pi^{(\mu_k)}_{k}$ as well as all previous basis operators. If we now assume that Eq.~\eqref{CLMarkovCond1} holds for each of our finite set of basis elements, i.e., the conditional state is the same for each outcome $\Pi^{(\mu_k)}_{k}$ and for each set of basis operations $\{\mathcal{A}_{j}^{{(\mu,\nu)}_{j}}\}$, we can take the state out of the sum:
\begin{align}
\rho_{l}(P_{k}^{(s)},\Pi^{(r)}_{k};\mathbf{A}_{k-1:0})=& \sum_{\vec{\mu},\vec{\nu},\mu_k} \alpha_{r,\mu_k}\alpha_{(\vec{\mu},\vec{\nu})} p_{l}(\mu_k,\vec{\mu},\vec{\nu}) \rho_l(P_{k}^{(s)}|\Pi^{(\mu_k)}_{k};\mathcal{A}_{k-1}^{{(\mu,\nu)}_{k-1}};\dots;\mathcal{A}_{1}^{{(\mu,\nu)}_{1}};\mathcal{A}_{0}^{{(\mu,\nu)}_0}) \nonumber \\
=& \sum_{\vec{\mu},\vec{\nu},\mu_k} \alpha_{r,\mu_k}\alpha_{({\vec{\mu}},\vec{\nu})} p_{l}(\mu_k,\vec{\mu},\vec{\nu}) \rho_l(P_{k}^{(s)}|\Pi^{(\mu_k')}_{k};\mathcal{A}_{k-1}^{{(\mu,\nu)}_{k-1}'};\dots;\mathcal{A}_{1}^{{(\mu,\nu)}_{1}'};\mathcal{A}_{0}^{{(\mu,\nu)}_0'}) \nonumber \\
=& \rho_l(P_{k}^{(s)}) \sum_{\vec{\mu},\vec{\nu},\mu_k} \alpha_{r,\mu_k}\alpha_{(\vec{\mu},\vec{\nu})} p_{l}(\mu_k,\vec{\mu},\vec{\nu}). \label{Markovbasis}
\end{align}

Since, by definition, the conditional state $\rho_l(P_{k}^{(s)})$ is a trace one object, it must be that 
\begin{equation}
\sum_{\vec{\mu},\vec{\nu},\mu_k} \alpha_{r,\mu_k}\alpha_{(\vec{\mu},\vec{\nu})} p_{l}(\mu_k,\vec{\mu},\vec{\nu}) = {\rm tr}[\rho_{l}(P_{k}^{(s)},\Pi^{(r)}_{k};\mathbf{A}_{k-1:0})] = p_{l}(\Pi^{(r)}_{k};\mathbf{A}_{k-1:0}).
\end{equation}
Dividing through by this quantity in Eq.~\eqref{Markovbasis}, we find an expression for the overall conditional state:
\begin{equation}
\rho_{l}(P_{k}^{(s)}|\Pi^{(r)}_{k};\mathbf{A}_{k-1:0})= \frac{\rho_{l}(P_{k}^{(s)},\Pi^{(r)}_{k};\mathbf{A}_{k-1:0})}{p_{l}(\Pi^{(r)}_{k};\mathbf{A}_{k-1:0})} = \rho_l(P_{k}^{(s)}),
\end{equation}
which is independent of the measurement outcome $\Pi^{(r)}_{k}$ and the past history of operations $\mathbf{A}_{k-1:0}$. Despite only assuming Eq.~\eqref{CLMarkovCond1} holds for a fixed set of inputs, we have shown that it holds for \emph{any} possible input prior to the causal break. Ergo, the process is Markovian.
\hfill $\blacksquare$

\section{Examples}\label{app:ex}

We have given a necessary and sufficient conditions for a quantum process to be Markovian. Here, using our formalism we present examples where various non-Markovianity witnesses fail to detect non-Markovian behaviour. The importance of these witnesses should be stressed: they enable efficient criteria to determine whether a process is non-Markovian in many cases.

\subsection{Divisibility}\label{app:ss}

\begin{figure}[ht]
\begin{center}
\includegraphics[width=.90\linewidth]
{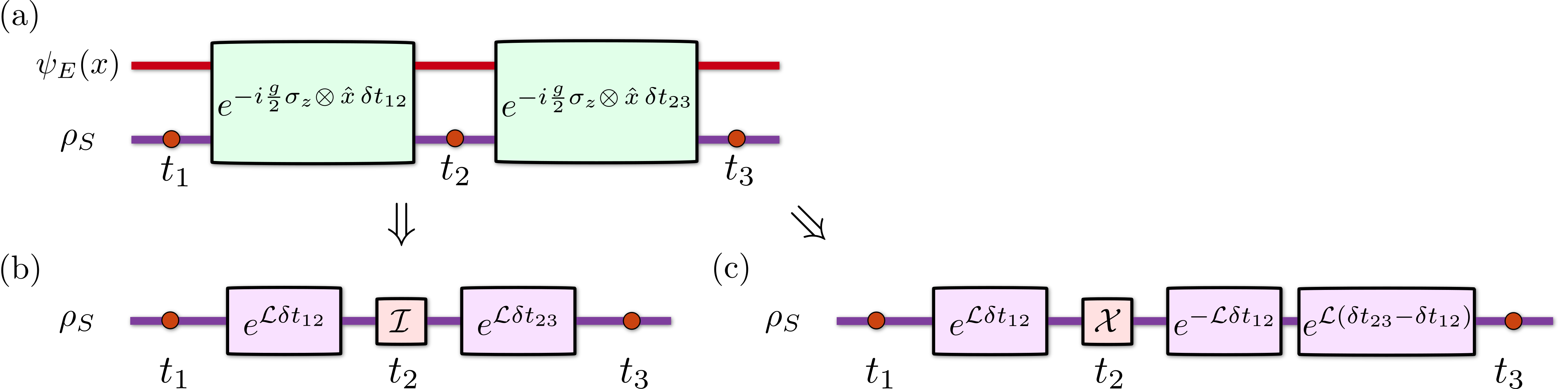}
\caption{\emph{A \textsc{cp}-divisible, but non-Markovian process.} (a) A qubit system in an arbitrary state $\rho_{S}$ evolves according to the Hamiltonian $H_{SE}= \frac{g}{2}\sigma_x \otimes \hat{x}$ along with an environmental position degree of freedom, which is initially uncorrelated with a Lorentzian wavefunction $\braket{x}{\psi}=\psi_E(x)=\sqrt{\gamma/\pi}/(x+i\gamma)$. (b) The reduced dynamics of the system is pure dephasing in the $\sigma_z$ basis, and can be written exactly in GKSL form, i.e., if the system is not interfered with, the evolution between any two points is a \textsc{cp}-map of the following form: $\rho(t_j) = \exp(\mathcal{L} \delta t_{ij})[\rho(t_i)]$, where $\delta t_{ij} = t_j-t_i$. It is therefore \textsc{cp}-divisible~\cite{sabri, NMrev, RevModPhys.88.021002}. (c) If an $\mathcal{X}$ operation ($\mathcal{X}[\rho]=\sigma_x\rho\sigma_x$) is performed at some time $t_2$, then the dynamics reverses for a period $\delta t_{12}$, such that the state at time $t_2+ \delta t_{12}$ is equal to the initial state $\rho_{S}$ up to a further $\mathcal{X}$ operation. The subsequent evolution is again pure dephasing. This behaviour constitutes a non-Markovian memory.
\label{image-cpdivisibility}}
\end{center}
\vspace{-20pt}
\end{figure}

Our first example is taken from Ref.~\cite{PhysRevA.92.022102}, and depicted here in Fig.~\ref{image-cpdivisibility}. The authors of Ref.~\cite{PhysRevA.92.022102} consider a qubit coupled to a continuous degree of freedom. They show that the exact dynamics of the qubit are fully \textsc{cp}-divisible, i.e., they are described by a time-independent generator $\mathcal{L}$ in Gorini-Kossakowski-Sudarshan-Lindblad (GKSL) form. This implies $\Lambda_{t:0} = \Lambda_{t:\tau} \circ \Lambda_{\tau:0}$ for any $\tau <t$ and all $\Lambda_{x:y}$ are \textsc{cptp}~maps. Under this evolution, the off-diagonal elements of the qubit decay exponentially in time (resulting from the entanglement growth between system and environment). However, it is shown that applying an $\mathcal{X}$ operation to the system at time $t_2>t_1$ and then at $2t_2-t_1$ fully returns the system to its state at $t_1$. Reversal of this exponential decay, which occurs for a time that depends on the system's history, implies that the dynamics are non-Markovian even according to, for example, the trace-distance distinguishability criterion discussed below. By introducing a causal break, it is also straightforward, if tedious, to show that it is also non-Markovian according our Theorem. This is an example of a process where the memory effects only appear in multi-time correlations.

However, in Ref.~\cite{PhysRevLett.101.150402} a \textsc{cptp}~map $\Lambda$ is defined to be Markovian if it can be written as $\Lambda=e^{\mathcal{L}}$. The motivation for this approach is to determine whether is $\Lambda$ is related to a valid generator for GKSL dynamics. As mentioned already, the example of Ref.~\cite{PhysRevA.92.022102} leads to dynamics of exactly this form, with positive and time-independent rate coefficients. Therefore, the snapshot approach would find this example to be Markovian. As we have argued, these dynamics are indeed non-Markovian, demonstrating the limitations of the snapshot method.
 
\subsection{Trace distance} \label{app:tc}

\begin{figure}[ht]
\begin{center}
\includegraphics[width=.95\linewidth]
{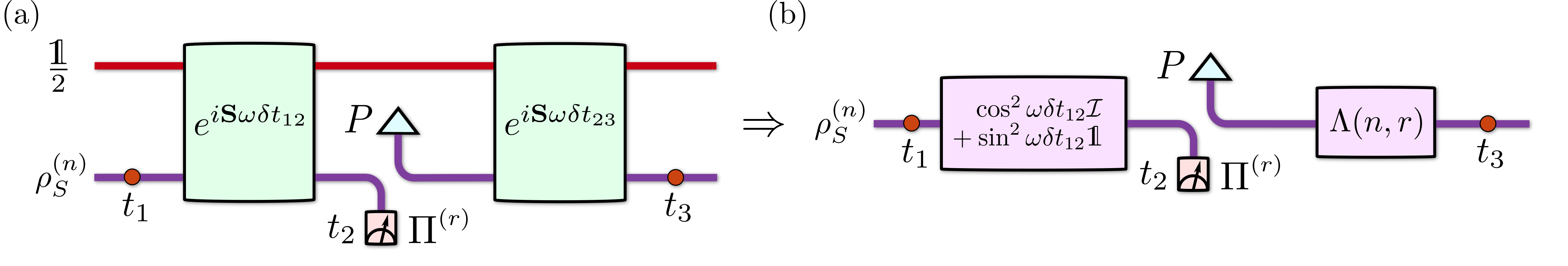}
\caption{\emph{A monotonically trace-distance decreasing, but non-Markovian process.} (a) System and environment (both qubits) evolve under a partial swap operation $U_{j:i}=\exp(i\mathbf{S} \omega \delta t_{ij}) = \cos\omega (t_j-t_i) \mathbbm{1}\otimes\mathbbm{1} + i\sin \omega (t_2-t_1) \mathbf{S}$. (b) If a measurement is made at some time $t_2$ and fresh pure state $P$ is prepared, then the subsequent reduced dynamics $\Lambda(n,r)$ depends on the measurement outcome $\Pi^{(r)}$ and the choice of initial state $\rho^{(n)}_{S}$ at time $t_1$. However, for $\omega(t_3-t_1)\leq \pi/2$, the process is monotonically trace-distance distinguishability decreasing. 
\label{image-tracedistanceexample}}
\end{center}
\vspace{-20pt}
\end{figure}

Consider the circuit presented in Fig.~\ref{image-tracedistanceexample}. The initial state of the system-environment at time $t_1$ is $\rho^{(n)}_{SE}(t_1) = \rho^{n}_{S} \otimes \mathbbm{1}/2$, where the initial system state is chosen from some fixed set, labelled by $n$.  After evolution under the partial swap operation $U_{2:1}=\exp(i\mathbf{S} \omega \delta t_{12})$, the total state at some later time $t_2$ is given by $\rho^{(n)}_{SE}(t_2) = \cos^2(\omega \delta t_{12}) \rho^{n}_{S} \otimes \mathbbm{1}/2 + \sin^2(\omega \delta t_{12}) \mathbbm{1}/2 \otimes \rho^{n}_{S} +i \cos(\omega \delta t_{12}) \sin(\omega \delta t_{12}) [\mathbf{S}, \rho^{n}_{S} \otimes \mathbbm{1}/2]$. The action on the system alone corresponds to depolarising channel $\Lambda_{2:1} = \cos^2(\omega \delta t_{12}) \mathcal{I} + \sin^2(\omega \delta t_{12}) \mathbbm{1}$, such that the state of the system at time $t_2$ is $\rho^{n}_{S}(t_2)=\cos^2(\omega\delta t_{12}) \rho^{n}_{S} + \sin^2(\omega\delta t_{12}) \mathbbm{1}/2$. 

Now suppose we initialise the system in two different states. The trace distance between these two states at a later time $t_3$ is 
\begin{equation}
{\rm tr} \left| \rho^{m}_{S}(t_2) -\rho^{n}_{S}(t_2) \right| = \cos^2(\omega \delta t_{12}) \;{\rm tr} \left| \rho^{m}_{S} -\rho^{n}_{S} \right|. 
\end{equation}
This is a monotonically decreasing function in the interval $\omega \delta t_{13} \in [0, \pi/2]$. Therefore in this interval the process will be labeled Markovian as determined by the measure proposed in Ref.~\cite{PhysRevLett.103.210401}.

However, consider a measurement on the system $\{\Pi^{(k)}\}$ followed by preparation in pure state $P$ at time $t_2$. The (normalised) total state after this causal break depends on the outcome of the measurement $\Pi^{(r)}$ and is given by ${\rho^{(n,r)}_{SE}}(t_2) = P \otimes \rho^{(n,r)}_{E}(t_2)$, where the operator on the environment is 
\begin{align}
\rho^{(n,r)}_{E}(t_2) =& 
\frac{{\rm tr}_{S}\left[\rho^{(n,r)}_{SE}(t_2) \Pi^{(r)} \right]}
{{\rm tr}\left[\rho^{(n,r)}_{SE}(t_2) \Pi^{(r)} \right]} \nonumber \\
=& \left({\rm tr}[\rho^{n}_{S}(t_2) \Pi^{(r)}] \cos^2(\omega \delta t_{12}) \;  \mathbbm{1} + {\rm tr}[\Pi^{(r)}]\sin^2(\omega \delta t_{12}) \; \rho^{n}_{S} \vphantom{+ i \cos(\omega\delta t_{12}) \sin(\omega\delta t_{12}) {\rm tr}_{S} \left[\Pi^{(r)} [\mathbf{S}, \rho^{n}_{S} \otimes  \mathbbm{1}]\right]}\right. \nonumber\\
&\quad\left.\vphantom{{\rm tr}[\rho^{n}_{S}(t_2) \Pi^{(r)}] \cos^2(\omega \delta t_{12}) \;  \mathbbm{1} + {\rm tr}[\Pi^{(r)}]\sin^2(\omega \delta t_{12}) \; \rho^{n}_{S}} + i \cos(\omega\delta t_{12}) \sin(\omega\delta t_{12}) {\rm tr}_{S} \left[\Pi^{(r)} [\mathbf{S}, \rho^{n}_{S} \otimes  \mathbbm{1}]\right]\right)\nonumber \\
&/\left(2{\rm tr}\left[\rho^{n}_{S}(t_2) \Pi^{(r)} \right] \cos^2(\omega \delta t_{12}) + {\rm tr}[\Pi^{(r)}]\sin^2(\omega \delta t_{12})\right).
\end{align}

The crucial point is that, after the causal break, there are no correlations with the environment, and the state of the system is reset to a pure state. Moreover, independent of the choice of the initial system state, i.e. $n$, the trace distance between the states of the system is zero after the fresh preparation. However, the environment state still depends on the initial state of the system (and the measurement outcome). If we let the evolution continue to some time $t_3$, the state of the system is 
\begin{align}
\rho^{n}_{S}(t_3)  =& \cos^2(\omega \delta t_{23}) P + \sin^2(\omega \delta t_{23}) \rho^{(n,r)}_{E}(t_2) +i  \cos(\omega\delta t_{23}) \sin(\omega \delta t_{23}) {\rm tr}_{S} \left([\mathbf{S}, P \otimes \rho^{(n,r)}_{E}(t_2)]\right) \nonumber \\
=& \Lambda(n,r)[P].
\end{align}
This state is a function of $\rho^{(n,r)}_{E}(t_2)$, which in turn is a function of the initial choice $n$ and measurement outcome $r$. Therefore, this process is operationally non-Markovian according to our main Theorem. For it to be operationally Markovian, the state of the system at $t_3$ (after the causal break) must only be a function of $P$, the system-environment unitary interaction, and the state of the environment, which \emph{cannot} be a function of past states of the system.

\subsection{Non-Markovianity without correlations}\label{app:cor}

\begin{figure}[ht]
\begin{center}
\includegraphics[width=.45\linewidth]
{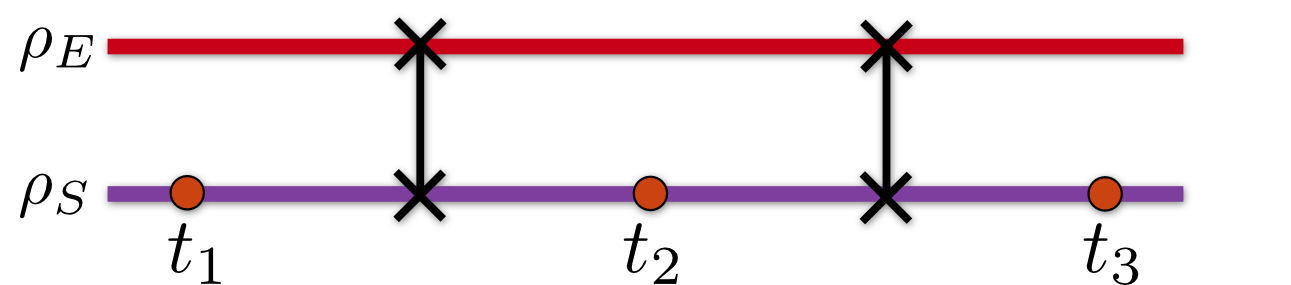}
\caption{\emph{A non-Markovian process without ${S\mbox{-}E}$ correlations.} The initial ${S\mbox{-}E}$ state is the product state $\rho_{S} \otimes \rho_{E}$, where system and environment have the same dimension. After time $t_1$ a \textsc{swap} operation is performed between them, such that the state at time $t_2$ is $\rho_{E} \otimes \rho_{S}$. Another \textsc{swap} operation is then performed. The joint ${S\mbox{-}E}$ state is always a product, i.e., there are never any ${S\mbox{-}E}$ correlations. However, the system state at $t_3$ is always $\rho_{S}$ (the state at $t_1$) independently of what operation is performed at $t_2$.
\label{image-correlation}}
\end{center}
\vspace{-20pt}
\end{figure}

As we mentioned earlier, when the initial ${S\mbox{-}E}$ correlations contribute to the dynamics of ${S}$, the process is non-Markovian. There are several witnesses for detecting initial ${S\mbox{-}E}$ correlations~\cite{mazzola2012dynamical, rodriguez2008completely, modidis, IC-breuer, rodriguez2012unification, PhysRevLett.107.180402, Gessner:2014kl}. However, correlations are not always important for quantum non-Markovian dynamics. Here we present a counter example.

Consider the two-step discrete process depicted in Fig.~\ref{image-correlation}, where both the ${S\mbox{-}E}$ unitaries are \textsc{swap} operations $\mathbf{S}$. We prepare any state $\rho_{S}$ initially for the system, and the initial state of the environment is $\rho_{E}$. After the first \textsc{swap} operation, the system will be in state $\rho_{E}$, independent of its initial state. While the environment will be in in the state $\rho_{S}$. Now, once again we can make any operation on the system we like (including a causal break) and allow for the second step in the process to take place. Independent of our preparation at the intermediary step, the system's state at the next step will be the same as the initial state of the system $\rho_{S}$. Since the reduced dynamics between times $t_2$ and $t_3$ clearly depends on the system state at earlier time $t_1$, the process is non-Markovian according to our Theorem. However, at no point in the process were there any correlations between the system and the environment. Note that if $\rho_{S}$ and $\rho_{E}$ are pure, then they also cannot be correlated to any third party.

\end{document}